\documentclass[aps, prb, twocolumn, amssymb, amsmath, showpacs, superscriptaddress]{revtex4-1}
\usepackage{bm}
\usepackage{times}
\usepackage{graphicx}
\usepackage{color}
\usepackage{dcolumn}
\usepackage[colorlinks=true, letterpaper=true, pdfstartview=FitV, linkcolor=blue, citecolor=blue, urlcolor=blue]{hyperref}
\usepackage{appendix}
\usepackage[normalem]{ulem}

\newcommand{\vect}[1]{\boldsymbol{#1}}                 %%% Used by Longjun
\usepackage{helvet}
\usepackage{pifont}
\newcommand{\cmark}{\ding{51}}
\newcommand{\xmark}{\ding{55}}
\usepackage{multirow}
\usepackage{siunitx}
\usepackage{makecell}

\usepackage{colortbl}
\usepackage{xcolor}
\definecolor{lightgreen}{rgb}{0.88, 1.0, 0.88}

\begin{document}

\title{Dissipationless Photovoltaic Spin Hall Effect from Spin-current Vorticity}
\author{Longjun Xiang}
\email[]{xianglj@szu.edu.cn}
\affiliation{College of Physics and Optoelectronic Engineering, Shenzhen University, Shenzhen 518060, China}
\author{Jian Wang}
\email[]{jianwang@hku.hk}
\affiliation{College of Physics and Optoelectronic Engineering, Shenzhen University, Shenzhen 518060, China}
\affiliation{Department of Physics, The University of Hong Kong, Pokfulam Road, Hong Kong, China}

\begin{abstract}
Spin-current vorticity (SCV) can generate the linear magnetic spin Hall effect [\href{https://doi.org/10.1038/s41586-018-0853-0}{Nature \textbf{565}, 627 (2019)}],
yet its role in nonlinear spin Hall transport has been much less explored.
Here, we show that, under a dc electric field,
SCV can deflect optically excited electrons to drive a dissipationless photovoltaic spin Hall effect (PSHE),
in which optical excitation by circularly and linearly polarized light is governed by the Berry curvature and quantum metric, respectively.
Because the Berry curvature is $\mathcal{T}$-odd whereas the quantum metric is $\mathcal{T}$-even,
their respective combinations with the $\mathcal{T}$-odd SCV give rise to $\mathcal{T}$-even and $\mathcal{T}$-odd PSHEs,
where $\mathcal{T}$ denotes time-reversal symmetry.
Remarkably, we find that the spin current of the $\mathcal{T}$-even PSHE can be reversed by switching the light helicity,
as illustrated in monolayer WTe$_2$.
By contrast, the $\mathcal{T}$-odd PSHE in altermagnets changes sign upon N\'eel-vector reversal,
as demonstrated in a $d$-wave altermagnetic model.
Beyond the PSHE, we show that the SCV dipole governs both the Drude and intrinsic nonlinear spin Hall effects proposed recently.
Our results reveal two switchable spin Hall mechanisms and establish SCV as a unifying concept for understanding dissipationless nonlinear spin Hall transport.
\end{abstract}
\maketitle

\bigskip
\noindent{\textcolor{blue}{\textit{Introduction}}}---The Berry curvature (BC)~\cite{Xiao2010} plays a pivotal role
in a variety of dissipationless charge-transport phenomena in quantum materials.
For example, in magnetic topological insulators, BC gives rise to the quantum anomalous Hall effect~\cite{TKNN1982, Haldane1988, Yu2010, Chang2013},
whereas in magnetic metals and Weyl semimetals, it drives the intrinsic anomalous Hall effect~\cite{Fang2003, Nagaosa2010, Burkov2014, Liu2018}.
In addition to these linear Hall responses,
momentum-space multipoles of the BC can generate nonlinear anomalous Hall effects.
In nonmagnetic Weyl semimetals and topological insulators,
the BC dipole can induce a second-order nonlinear anomalous Hall effect~\cite{Sodemann2015, Ma2019, Kang2019}.
Furthermore, in the emergent altermagnets~\cite{AM1, AM2, AM3, AM4, AM41, AM5, AM6, AM7, AM8, AM9},
the BC quadrupole~\cite{KTLawPRB, BCQ2, BCQ3, BCQ4, BCQ5, Xiang2026} can drive a third-order nonlinear anomalous Hall effect.

Beyond charge Hall transport,
the spin degree of freedom in quantum materials can further
support the dissipationless spin Hall effect~\cite{Sinova2015RMP,Sinova2004PRL,Murakami2003Science,spinBC},
including contributions arising from the $SU(2)$ non-Abelian holonomy~\cite{Murakami2003Science} and the spin Berry curvature~\cite{spinBC}.
These dissipationless spin responses provide an important route toward energy-efficient spintronic applications~\cite{Zutic2004RMP}.
In recent years, the magnetic spin Hall effect~\cite{Kimata2019, Holanda2020, Mook2020, Hu2022, Dai2024, Xiang2025SHE, Tao2026},
first experimentally observed in noncollinear antiferromagnets~\cite{Kimata2019},
has attracted considerable attention.
It has been shown to originate from the spin-current vorticity (SCV)~\cite{Mook2020},
which, analogous to BC, generates a dissipationless transverse spin response.
Despite this progress, how SCV governs nonlinear spin Hall transport remains largely unexplored,
in stark contrast to the central role of BC in nonlinear charge Hall effects.

\begin{figure}[t!]
\includegraphics[width=0.90\columnwidth]{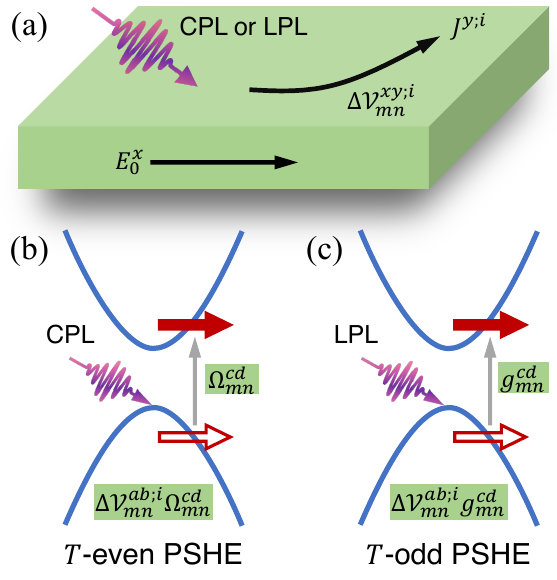}
\caption{(a) The photovoltaic spin Hall effect (PSHE) driven by a dynamical optical field together with a static electric field.
(b) The $\mathcal{T}$-even PSHE resulting from the interplay between the differential spin-current vorticity
$\Delta\mathcal{V}_{nm}^{ab;i}=\mathcal{V}_{nn}^{ab;i}-\mathcal{V}_{mm}^{ab;i}$ and the Berry curvature $\Omega_{nm}^{cd}$.
(c) The $\mathcal{T}$-odd PSHE resulting from the interplay between the $\Delta\mathcal{V}_{nm}^{ab;i}$ and the quantum metric $g^{cd}_{nm}$.}
\label{FIG1}
\end{figure}

In this Letter, we propose that the SCV, in a static electric field,
can deflect the optically excited Bloch electrons in both metallic and insulating quantum materials to generate a dissipationless transverse spin current,
leading to a novel photovoltaic spin Hall effect (PSHE).
The optical excitation under circularly (linearly) polarized light
is governed by the $\mathcal{T}$-odd Berry curvature ($\mathcal{T}$-even quantum metric),
which combines with the $\mathcal{T}$-odd SCV to yield a $\mathcal{T}$-even ($\mathcal{T}$-odd) PSHE,
where $\mathcal{T}$ denotes time-reversal symmetry.
Notably, we show that the spin Hall current of the $\mathcal{T}$-even PSHE can be reversed by switching the light helicity,
as illustrated in monolayer WTe$_2$.
Furthermore, the $\mathcal{T}$-odd PSHE can be used to detect the reversal of N\'eel vector in altermagnets,
as illustrated in a $d$-wave altermagnetic model.
In addition, we show that the SCV dipole governs both the Drude~\cite{Xiang2025SHE}
and intrinsic~\cite{Zhang2024, Wang2025, Sarkar2026} nonlinear spin Hall effects proposed recently.
Our work reveals two switchable spin Hall mechanisms,
which may be applicable to opto-spintronics~\cite{Nemec2018AFOptospin, Sierra2021, Gish2024},
and establishes SCV as a unifying concept for understanding dissipationless nonlinear spin Hall transport.

\bigskip
\noindent \textcolor{blue}{\noindent{\textit{Quantum Theory of PSHE}}}---Within the independent-particle approximation,
the spin current density $J^{a;i}$ in crystalline solids can be evaluated as
\begin{align}
J^{a;i}
\equiv
\sum_{mn}
\int_k
v^{a;i}_{mn}
\rho_{nm},
\label{current}
\end{align}
where $\int_k \equiv \int d\vect{k}/(2\pi)^d$, with $d$ the spatial dimension,
and $v^{a;i}_{mn}$ and $\rho_{nm}$ are the matrix elements of the spin-current operator
$\hat{v}^{a;i} \equiv (\hat{v}^a \hat{s}^i+\hat{s}^i \hat{v}^a)/2$ and the density matrix $\hat{\rho}$ in the Bloch basis $|n\rangle$, respectively.
Here, $\hat{v}$ is the velocity operator and $\hat{s}^i=\hat{\sigma}^i/2$ is the spin angular momentum operator,
where $\hat{\sigma}^i$ is the Pauli matrix for spin, with $i=\{x, y, z\}$.
The density matrix element $\rho_{nm}$ obeys the quantum Liouville equation,
which, in the presence of an electric field, is given by~\cite{Sipe1995, Sipe2000}
\begin{align}
i \partial_t \rho_{nm}
=
\epsilon_{nm} \rho_{nm} &+ \sum_\alpha i \mathcal{D}_{nm}^b\rho_{nm} E_\alpha^b(t)
\nonumber \\
&+
\sum_{\alpha,l} \left( r^b_{nl}\rho_{lm} - \rho_{nl}r^b_{lm} \right) E_\alpha^b (t).
\label{fullEQ}
\end{align}
Here, $\epsilon_{nm}=\epsilon_{n}-\epsilon_{m}$ is the band energy difference,
$\mathcal{D}_{nm}^b \equiv \partial_b - i(\mathcal{A}_n^b-\mathcal{A}_m^b)$ is the $U(1)$-covariant derivative,
where $\partial_b \equiv \partial/\partial k_b$ and $\mathcal{A}_n^b$ is the intraband Berry connection,
$r_{nl}^b$ the interband Berry connection,
and $E^b_\alpha (t) = E_\alpha^b e^{-i(\omega_\alpha+i\eta)t}$ is the applied electric field,
where $\eta \rightarrow 0^+$ and $E^b_\alpha=E^b(\omega_\alpha)$, with $\omega_\alpha = \pm \omega$.
We assume $e=\hbar=1$ unless otherwise stated.

To derive the spin current driven by a dynamical optical field together with a static electric field,
we solve Eq.~(\ref{fullEQ}) iteratively to evaluate the third-order density matrix element $\rho_{nm}^{(3)} \propto (E^b_\alpha)^3$,
from which we extract the following component~\cite{sup}:
\begin{align}
\rho_{nn}^{(3)}
&=
\sum_{\alpha\beta,l}
i\partial_b
\left(
\dfrac{r^c_{nl}r^d_{ln}f_{nl}}
{\omega_\alpha-\epsilon_{ln}+i\eta}
\right)
\dfrac{E_0^b E_\beta^c E_\alpha^d e^{-i \Omega_1 t}}{\Omega_1^2}
\nonumber \\
&-
\sum_{\alpha\beta,l}
i\partial_b
\left(
\dfrac{r^d_{nl}r^c_{ln}f_{ln}}
{\omega_\alpha-\epsilon_{nl}+i\eta} 
\right)
\dfrac{E_0^b E_\beta^c E_\alpha^d e^{-i\Omega_1 t}}{\Omega_1^2}
\label{rho3}.
\end{align}
Here, $E_0^b$ is the static electric field, $\Omega_1=\omega_\alpha+\omega_\beta$,
and $f_{nl}=f_n-f_l$, with $f_n$ being the equilibrium Fermi distribution function.
Substituting Eq.~(\ref{rho3}) into Eq.~(\ref{current}) and taking the second time derivative of $J^{a; i}$,
we obtain the following resonant contribution to $\partial_t^2 J^{a;i}$~\cite{sup}
\begin{align}
\partial^2_t J^{a; i}
&=
\sum_{\alpha,nm}
\int_k f_{nm} \mathcal{K}_{nm}^{abcd;i}\delta(\omega_\alpha-\epsilon_{mn})E^b_0 E^{c*}_{\alpha} E^d_{\alpha},
\label{Ja30}
\end{align}
where $\mathcal{K}^{abcd;i}_{nm} \equiv i I^{ab;i}_{nm}G^{cd}_{nm}$,
with $I^{ab;i}_{nm} = \partial_b \left( v^{a;i}_{nn}-v^{a;i}_{mm} \right)$
and $G^{cd}_{nm} = r^c_{nm} r^d_{mn}$ being the quantum geometric tensor~\cite{NP2021}.
To derive Eq.~\eqref{Ja30}, we have used $\Omega_1=\omega_\alpha+\omega_\beta=0$.
We note that suppressing the spin degree of freedom by recasting $I^{ab;i}_{nm}$ into $I^{ab}_{nm} \equiv \partial_b(v_{nn}^a-v_{mm}^a)$,
Eq.~(\ref{Ja30}) reproduces the jerk photovoltaic effect~\cite{Fregoso,corrected0, corrected}.
Since $I^{ab}_{nm}=I^{ba}_{mn}$ so that $\partial_t^2 J^{a}E_0^a \neq 0$ in general.
As a result, the jerk charge current does not include a Hall component~\cite{NP2021} and is dissipative.
Without this suppression, Eq.~\eqref{Ja30} gives rise to the jerk spin photovoltaic effect~\cite{sup},
which contains both the longitudinal dissipative and the transverse dissipationless contributions due to $I^{ab;i}_{nm} \neq I^{ba;i}_{nm}$.

\begin{center}
\begin{table}
\caption{\label{tab1} 
The parity of the differential SCV $\Delta\mathcal{V}_{mn}^{ab;i}$, the Berry curvature $\Omega^{cd}_{mn}$,
the quantum metric $g^{cd}_{mn}$, and $\kappa^{abcd;i}_{L/C}$ determined by them,
under the $\mathcal{T}$-symmetry, $\mathcal{P}$-symmetry, and the combined $\mathcal{P}\mathcal{T}$-symmetry.
Here~\cmark~and~\xmark~denote even and odd parities, respectively.
}
\begin{tabular}{c|c|c|c|c|c}
\hline
\hline
                             & \ \ \ $\Delta\mathcal{V}_{nm}^{ba;i}$ \ \ \  & \ \ \  $g^{cd}_{nm}$ \ \ \  & \ \ \  $\Omega^{cd}_{nm}$ \ \ \
                             & \ \ \ $\kappa^{abcd;i}_{L}$ \ \ \ & \ \ \ $\kappa^{abcd;i}_{C}$ \ \ \ \\
\hline
\ \ \ $\mathcal{T}$ \ \ \    & \ \ \ \xmark \ \ \  & \ \ \ \cmark \ \ \ & \ \ \ \xmark \ \ \ & \ \ \ \xmark \ \ \ & \ \ \ \cmark \ \ \ \\
\hline                                            
\ \ \ $\mathcal{P}$ \ \ \    & \ \ \ \cmark \ \ \  & \ \ \ \cmark \ \ \ & \ \ \ \cmark \ \ \ & \ \ \ \cmark \ \ \ & \ \ \ \cmark \ \ \ \\
\hline                                            
\ \ \ $\mathcal{P}\mathcal{T}$ \ \ \  & \ \ \  \xmark \ \ \  & \ \ \ \cmark \ \ \ & \ \ \ \xmark \ \ \ & \ \ \ \xmark \ \ \ & \ \ \ \cmark \ \ \ \\
\hline
\hline
\end{tabular}
\end{table}
\end{center}

To obtain its dissipationless contribution with $\partial_t^2 j^{a;i}E_0^a = 0$,
namely the PSHE, we antisymmetrize $a$ and $b$ of $I^{ab;i}_{nm}$:
\begin{align}
\frac{I^{ab;i}_{nm} - I^{ba;i}_{nm}}{2}
=
\dfrac{-\mathcal{V}^{ab;i}_{nn}+\mathcal{V}_{mm}^{ab;i}}{2}
\equiv
\dfrac{\Delta\mathcal{V}^{ab;i}_{mn}}{2},
\label{SCV}
\end{align}
where $\mathcal{V}^{ab;i}_{nn} = \partial_a v_{nn}^{b;i}-\partial_b v_{nn}^{a;i}$ is the spin-current vorticity~\cite{Mook2020} (SCV),
which is known as the origin of the magnetic spin Hall effect~\cite{Kimata2019} in magnetic metals,
and $\Delta\mathcal{V}^{ab;i}_{mn}=\mathcal{V}^{ab;i}_{mm}-\mathcal{V}^{ab;i}_{nn}$ is the differential SCV.
Inserting Eq.~\eqref{SCV} into Eq.~\eqref{Ja30} and defining 
$\partial_t^2 J^{a;i} \equiv
2\kappa_{L}^{abcd;i} E^b_0 \text{Re}[E^c_{\omega} E^{d*}_{\omega}]+2\kappa_{C}^{abcd;i} E^b_0 \text{Im}[E^c_{\omega} E^{d*}_{\omega}]$,
where $E^c_\omega \equiv E^c(\omega)$ and $E^{d*}_\omega \equiv E^d(-\omega)$,
the response tensors for PSHE~\cite{sup} are found to be:
\begin{align}
\kappa_{L}^{abcd;i} &= \dfrac{\pi e^3}{2\hbar^2} \sum_{nm}\int_k f_{nm}  \Delta\mathcal{V}_{mn}^{ab;i} g^{cd}_{mn}
\delta(\omega-\epsilon_{mn}),
\label{kappaL}
\\
\kappa_{C}^{abcd;i} &= \dfrac{\pi e^3}{4\hbar^2} \sum_{nm}\int_k f_{nm}  \Delta\mathcal{V}_{mn}^{ba;i} \Omega^{cd}_{mn}
\delta(\omega-\epsilon_{mn}),
\label{kappaC}
\end{align}
where $g^{cd}_{mn}=g^{cd}_{nm}\equiv \text{Re}G^{cd}_{nm}$ is the quantum metric
and $\Omega^{cd}_{mn}=-\Omega^{cd}_{nm}\equiv2\text{Im}G^{cd}_{nm}$ is the Berry curvature.
Here, $e$ and $\hbar$ are restored by dimensional analysis.
As indicated by their subscripts $L$ and $C$,
$\kappa_{L}^{abcd;i}$ can be excited by the linearly polarized light (LPL),
which ensures that $\text{Re}[E^c_{\omega} E^{d*}_{\omega}] \neq 0$,
while $\kappa_{C}^{abcd;i}$ can be activated by the circularly polarized light (CPL),
which guarantees that $\text{Im}[E^c_{\omega}E^{d*}_{\omega}] \neq 0$.
Notably, since $\text{Re}[E^c_{\omega} E^{d*}_{\omega}]=\text{Re}[E^d_{\omega} E^{c*}_{\omega}]$
and $\text{Im}[E^c_{\omega}E^{d*}_{\omega}]=-\text{Im}[E^d_{\omega}E^{c*}_{\omega}]$,
the optical excitations of $\kappa_L^{abcd;i}$ and $\kappa_C^{abcd;i}$, respectively,
are controlled by the symmetric quantum metric $g^{cd}_{nm}$ and the antisymmetric Berry curvature $\Omega^{cd}_{nm}$,
as illustrated in Fig.~\ref{FIG1}(b) and \ref{FIG1}(c).
After completing the excitations, the differential SCV $\Delta\mathcal{V}^{ab;i}_{mn}$ further deflects
the excited electrons (carrying a spin $\hat{\sigma}^i$) to generate the PSHE, as illustrated in Fig.~\ref{FIG1}(a).

Equations~(\ref{kappaL}) and (\ref{kappaC}) are the main results of this Letter.
Several remarks are in order.
First, both $\kappa_L^{abcd;i}$ and $\kappa_C^{abcd;i}$ are gauge invariant, since
$g_{mn}^{cd}$, $\Omega_{mn}^{cd}$, and $\Delta \mathcal{V}_{mn}^{ba;i}$ are
invariant under the $U(1)$ gauge transformation $|n\rangle\rightarrow e^{i\phi_n}|n\rangle$.
Therefore, these expressions can be combined with state-of-the-art first-principles calculations
to identify the candidate materials for the PSHE, in analogy with shift and injection currents~\cite{Wanghua2020}.
Second, as listed in Table~\ref{tab1}, $\Delta \mathcal{V}_{mn}^{ba;i}$ and $\Omega_{mn}^{cd}$ are $\mathcal{T}$-odd,
whereas $g_{mn}^{cd}$ is $\mathcal{T}$-even;
as a result, $\kappa_C^{abcd;i}$, determined by $\Delta \mathcal{V}_{mn}^{ba;i}\Omega_{mn}^{cd}$,
gives rise to a $\mathcal{T}$-even PSHE,
while $\kappa_L^{abcd;i}$, determined by $\Delta \mathcal{V}_{mn}^{ba;i}g_{mn}^{cd}$, leads to a $\mathcal{T}$-odd PSHE.
Furthermore, since $\Delta \mathcal{V}_{mn}^{ba;i}$, $\Omega_{mn}^{cd}$, and $g_{mn}^{cd}$ are all $\mathcal{P}$-even,
both $\kappa_L^{abcd;i}$ and $\kappa_C^{abcd;i}$ are $\mathcal{P}$-even,
which indicates that the PSHE can generally appear in centrosymmetric solids.
To close this section, we remark that $\kappa_C^{abcd;i}$ driven by CPL is helicity dependent,
enabling the reversal of the spin Hall current by switching the light helicity,
just as the circular injection current~\cite{helicity}.
In addition, the spin Hall current given by $\kappa_L^{abcd;i}$ in altermagnets
can also be reversed but by switching the N\'eel order,
considering that $\kappa_L^{abcd;i}$ is also $\mathcal{PT}$-odd (see Table~\ref{tab1}).
Below we illustrate these general features of PSHE.

\begin{figure}[t!]
\includegraphics[width=0.90\columnwidth]{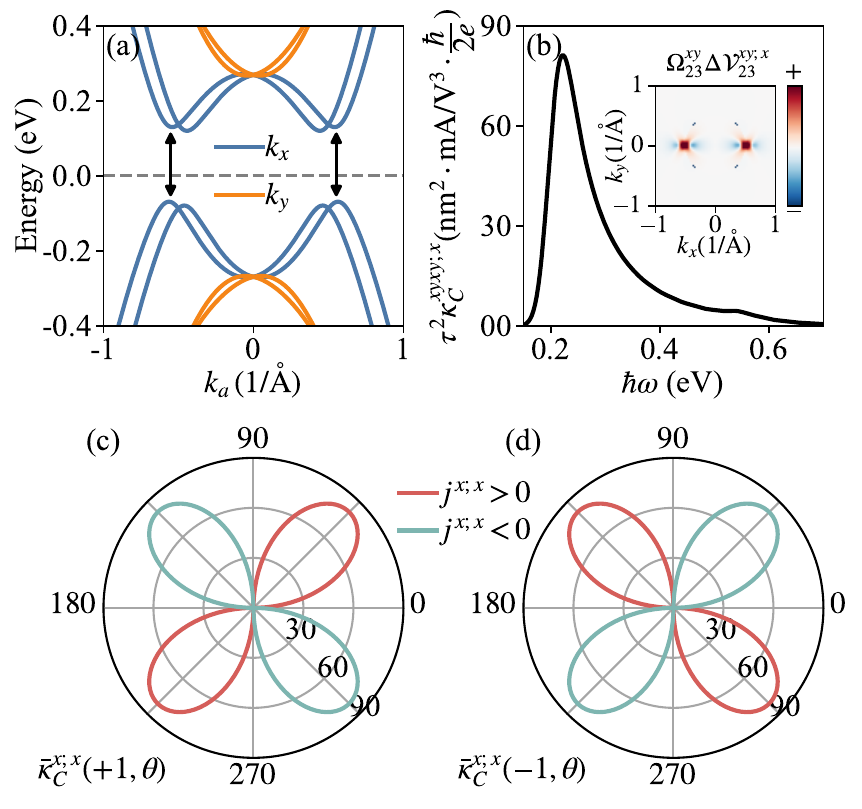}
\caption{(a) Band dispersion of monolayer WTe$_2$.
(b) Frequency dependence of $\kappa_C^{xyxy;x}$ for the $\mathcal{T}$-even PSHE.
The inset shows the $\vect{k}$-space distribution of the dominant integral kernel $\Omega^{xy}_{32}\Delta\mathcal{V}^{xy;x}_{32}$ of $\kappa_C^{xyxy;x}$.
(c-d) Angular dependence of the $\mathcal{T}$-even PSHE with $\chi=+1$ and $\chi=-1$, respectively,
where the positive and negative spin Hall currents are interchanged upon reversing the light helicity. 
Parameters: $A=0.1\mathrm{eV}\cdot\mathrm{\mathring{A}}^2$, $B=1 \mathrm{eV}\cdot\mathrm{\mathring{A}}^2$, $\delta=-0.25 \mathrm{eV}$, $D=0.1 \mathrm{eV}$,
$v_y=1 \mathrm{eV}\cdot\mathrm{\mathring{A}}$, $\tau=10^{-13}\,\mathrm{s}$, $\mu=0\,\mathrm{eV}$, and $\alpha_x=\alpha_y=0.1\,\mathrm{eV}\cdot\mathrm{\mathring{A}}$.
}
\label{FIG2}
\end{figure}

\bigskip
\noindent{\textcolor{blue}{\textit{Time-reversal-even PSHE in monolayer WTe$_2$}}}---To demonstrate
the PSHE described by the $\mathcal{T}$-even $\kappa_C^{abcd;i}$,
we consider a two-dimensional (2D) system with $\mathcal{T}$ symmetry.
In this limit, we note that the spatial indices $a, b, c, d \in \{x, y\}$ while the spin index $i \in \{x, y, z\}$.
Furthermore, considering the antisymmetric permutation symmetry of $\kappa^{abcd;i}_C$,
namely $\kappa^{abcd;i}_C=-\kappa^{bacd;i}_C=-\kappa^{abdc;i}_C$,
the nonzero elements for the $\mathcal{T}$-even PSHE in 2D limit are
$\kappa^{xyxy;i}_C=-\kappa^{yxxy;i}_C=-\kappa^{xyyx;i}_C=\kappa^{yxyx;i}_C$.
We then concentrate on the independent component $\kappa^{xyxy;i}_C$ in the following.
Under mirror symmetry $\mathcal{M}_a$, the symmetry transformaton dictates that
$\kappa^{xyxy;i}_C=\det(\mathcal{M}_a)\mathcal{M}_a^{xx}\mathcal{M}_a^{yy}\mathcal{M}_a^{xx}\mathcal{M}_a^{yy}\mathcal{M}_a^{ii}\kappa^{xyxy;i}_C
=-\mathcal{M}_a^{ii}\kappa^{xyxy;i}_C$, which fixes $i=a$. 
Note that $\mathcal{M}_a$ with $a \in \{x, y\}$ is diagonal in its matrix representation.
On the other hand, under rotation symmetry $C_{nz}$ with $n=\{2,3,4,6\}$,
the symmetry transformaton dictates that
$\kappa^{xyxy;i}_C=\det(R)R^{xa}R^{yb}R^{xc}R^{yd}R^{ii'}\kappa^{abcd;i'}_C=R^{xa}R^{yb}R^{xc}R^{yd}R^{ii'}\kappa^{abcd;i'}_C$,
where $R^{ab}$ is the 2D matrix representation of $C_{nz}$. Note that $\det(R)=1$.
Thanks to the antisymmetric permutation symmetry of $\kappa_C^{abcd;i}$,
by defining $\kappa^{abcd;i}_C=\epsilon^{ab}\epsilon^{cd}K^i$, where $K^i=\kappa_C^{xyxy;i}$ and $\epsilon^{ij}$ is the 2D Levi-Civita symbol,
we obtain $K^i=(R^{xa}\epsilon_{ab}R^{yb})(R^{xc}\epsilon_{cd}R^{yd})R^{ii'}K^{i'}=R^{ii'}K^{i'}$.
Here, $R^{xa}\epsilon_{ab}R^{yb}=\det(R)=1$ and $R^{xc}\epsilon_{cd}R^{yd}=\det(R)=1$ have been used.
This equation has only a trivial solution since $\det(1-R)=4\sin^2(\pi/n)\neq0$ for $i, i' \in \{x, y\}$.
As a result, under $C_{nz}$, $\kappa_C^{xyxy;x}=\kappa_C^{xyxy;y}=0$ and the allowed element becomes $\kappa_C^{xyxy;z}$.

Owing to the stringent constraints imposed by rotational and mirror symmetries in 2D limit,
we choose the low-symmetry monolayer WTe$_2$ as a representative platform to illustrate the helicity-dependent PSHE.
Its spinful low-energy effective Hamiltonian can be written as
$H=H_0\hat{\sigma}_0+\alpha_xk_x\hat{\tau}_z\hat{\sigma}_y+\alpha_yk_y\hat{\tau}_z\hat{\sigma}_x$,
where $H_0=d_0+\vect{d}\cdot\hat{\vect{\tau}}$ with~\cite{WTe21,WTe22} $d_0=Ak^2$, $d_x=D$, $d_y=v_yk_y$, and $d_z=Bk^2+\delta$,
where $k^2 \equiv k_x^2+k_y^2$.
Here, $\hat{\tau}_i$ represents a pseudospin and the spinless Hamiltonian $H_0$ preserves $\mathcal{T}$ and $\mathcal{M}_x$ symmetries.
In order to calculate the PSHE, we have added two symmetry-allowed terms to make the Hamiltonian spinful,
whose band structure is shown in Fig.~\ref{FIG2}(a).
For this model, the $\mathcal{T}$-odd PSHE is identically forbidden while the allowed elements for the $\mathcal{T}$-even PSHE are
$\kappa^{xyxy;x}_C=-\kappa^{yxxy;x}_C=-\kappa^{xyyx;x}_C=\kappa^{yxyx;x}_C$,
which is determined by $\Delta\mathcal{V}^{xy;x}_{mn}\Omega^{xy}_{mn}$ in terms of Eq.~\eqref{kappaC}.
The dependence of $\kappa_C^{xyxy;x}$ on the photon frequency is shown in Fig.~\ref{FIG2}(b).
We observe a resonant peak near $\hbar\omega=0.22\,\mathrm{eV}$,
which mainly arises from the states near the minimal direct-gap region, as indicated by the vertical arrow in Fig.~\ref{FIG2}(a).
This can further be seen from the $\vect{k}$-space distribution of the integral kernel $\Omega^{xy}_{23}\Delta\mathcal{V}^{xy;y}_{23}$,
as shown in the inset of Fig.~\ref{FIG2}(b).

We remark that $\kappa_C^{xyxy;x}$ excited by CPL can give rise to a helicity-dependent spin Hall current.
To see this explicitly, we consider a monochromatic optical field with a general polarization angle $\theta$:
$\vect{E}(t) = E_1 \left( \cos\theta,\, i\chi\sin\theta,\, 0 \right) e^{-i\omega t} + c.c.$,
where $\chi = \pm 1$ denotes the helicity of light and $c.c.$ the complex conjugate of the first term.
The helicity-dependent drive is given by $2\text{Im}[ E^x_{\omega} E^{y*}_{-\omega} ] = \chi \sin2\theta E_1^2$
and the resulting spin Hall current takes the form
$j^{x;x} = \tau^2 \kappa_C^{xyxy;x} \chi \sin2 \theta E_0E_1^2 = \bar{\kappa}^{x;x}_C({\chi},\theta) E_0E_1^2$,
where $\bar{\kappa}_C^{x;x}(\chi,\theta) \equiv \tau^2\kappa_C^{xyxy;x} \chi \sin2 \theta$.
The angular dependence of $\bar{\kappa}^{x;x}_C({\chi},\theta)$ for $\chi=1$ and $\chi=-1$ are shown in Fig.~\ref{FIG2}(c) and ~\ref{FIG2}(d), respectively,
where we have taken $\tau^2\kappa^{xyxy;x}_C=80\,\mathrm{nm}^2\cdot\mathrm{mA}/\mathrm{V}^3\cdot\frac{\hbar}{2e}$ in terms of Fig.~\ref{FIG2}(b).
The helicity dependence of $j^{x;x}$ is manifested by the interchange of the positive and negative regions when the light helicity is reversed,
demonstrating that the spin Hall current of the $\mathcal{T}$-even PSHE satisfies
\begin{align}
j^{x;x}(\chi=1, \theta)=-j^{x;x}(\chi=-1, \theta).
\end{align}
Note that $j^{x;x}(\chi,\theta)=\bar{\kappa}^{x;x}(\chi,\theta)E_0E_1^2$ reaches its maximum magnitude for CPL,
where $\theta=(2n+1)\pi/4$ with $n=0, 1, 2, \cdots$,
while vanishes for LPL, where $\theta=n\pi/2$.

\begin{figure}[t!]
\includegraphics[width=0.90\columnwidth]{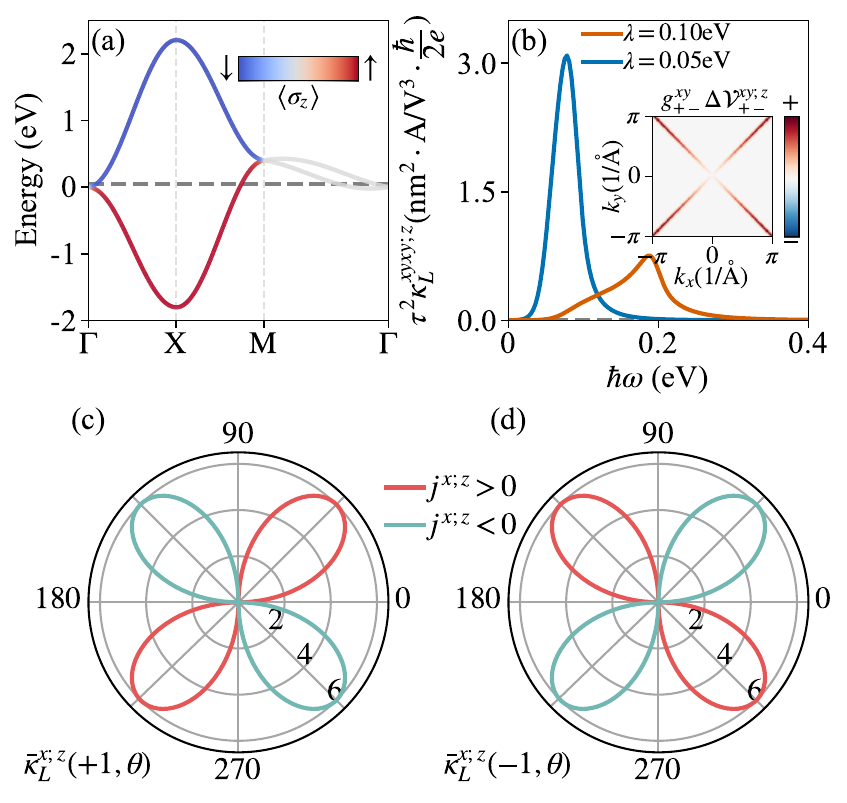}
\caption{(a) Band dispersions of $d$-wave altermagnet.
(b) Frequency dependence of the response tensor $\kappa_L^{xyxy;z}$ for the $\mathcal{T}$-odd PSHE.
The inset shows the $\vect{k}$-space distribution of the dominant integral kernel $g^{xy}_{+-}\Delta\mathcal{V}^{xy;z}_{+-}$ of $\kappa_L^{xyxy;z}$.
(c-d) Angular dependence of the $\mathcal{T}$-odd PSHE with $J=+1\,\mathrm{eV}$ and $J=-1\,\mathrm{eV}$, respectively,
where the positive and negative spin Hall currents are interchanged upon reversing the altermagnetic N\'eel order. 
Parameters: $|J|=1\,\mathrm{eV}$, $t=0.05\,\mathrm{eV}$, $\lambda=0.1\,\mathrm{eV}$,
$\tau=10^{-13}\,\mathrm{s}$, $T=100\,\mathrm{K}$, $\mu=0.05\,\mathrm{eV}$, and $a=1\,\mathring{\mathrm{A}}$.
}
\label{FIG3}
\end{figure}

\bigskip
\noindent{\textcolor{blue}{\textit{Time-reversal-odd PSHE in $d$-wave altermagnet}}}---To demonstrate the $\mathcal{T}$-odd PSHE,
we consider a $d$-wave altermagnet~\cite{AM2} described by the two-band tight-binding model $H=d_0+\vect{d}\cdot\hat{\vect{\sigma}}$,
where $d_0=2t(2-\cos a k_x - \cos a k_y)$, $d_x=\lambda \sin [a (k_x+k_y)/2]$, $d_y=\lambda \sin [a (k_y-k_x)/2]$, and $d_z=J(\cos a k_x-\cos a k_y)$.
Here, $J$ denotes the $d_{x^2-y^2}$ altermagnetic order parameter and
$\lambda$ incorporates the spin-orbit coupling (SOC). In the absence of SOC, namely $\lambda=0$,
this model conserves $\hat{\sigma}_z$ and can be described by a spin point group $^24/^1m^1m^2m$.
This indicates that the opposite-spin magnetic sublattices can be related by the nontrivial spin group operations
such as $[C_{2\perp}^s||C_4]$ and $[C_{2\perp}^s||\mathcal{M}_{x+y}]$,
where $C_{2\perp}^s$ denotes the twofold rotation in spin space about an axis perpendicular to the spin direction,
while $C_4$ and $\mathcal{M}_{x+y}$ denote, respectively, the fourfold rotation around the $z$ axis
and the mirror operation with respect to the $x=y$ plane in real space.
Once SOC is included, even if weak with $J \gg \lambda$,
the spin point group is reduced to the magnetic point group $4'/mm'm$,
which can be generated by $C_{4z}\mathcal{T}$, $\mathcal{M}_z$, and $\mathcal{M}_x\mathcal{T}$.

Due to the permutation symmetry $\kappa_{L}^{abcd;i}=-\kappa^{bacd;i}_L$ and $\kappa_{L}^{abcd;i}=\kappa^{abdc;i}_L$,
where $a, b, c, d \in \{x, y \}$ while $i \in \{x, y, z\}$ for this 2D system,
the allowed components are $\kappa_L^{xycd;i}=\kappa_L^{xydc;i}=-\kappa^{yxcd;i}=-\kappa^{yxdc;i}$.
Under $\mathcal{M}_z=\mathrm{diag}(1,1,-1)$,
the symmetry transformation gives $\kappa_L^{xycd;i} = \det(\mathcal{M}_z) \mathcal{M}_{z}^{xx} \mathcal{M}_{z}^{yy} \mathcal{M}_{z}^{cc} \mathcal{M}_{z}^{dd}
\mathcal{M}_{z}^{ii} \kappa_L^{xycd;i}=-\mathcal{M}_{z}^{ii} \kappa_L^{xycd;i}$.
Here we have used $\det(\mathcal{M}_z)=-1$.
As a result, the $i=x,y$ components are forbidden, while $i=z$ component can be allowed.
Furthermore, under $\mathcal{M}_x\mathcal{T}$ with $\mathcal{M}_x=\mathrm{diag}(-1,1,1)$,
the symmetry transformation gives
$\kappa_L^{xycd;z} = - \det(\mathcal{M}_x) \mathcal{M}_{x}^{xx} \mathcal{M}_{x}^{yy} \mathcal{M}_{x}^{cc} \mathcal{M}_{x}^{dd} \mathcal{M}_{x}^{zz}
\kappa_L^{xycd;z}=-\mathcal{M}_{x}^{cc} \mathcal{M}_{x}^{dd} \kappa_L^{xycd;z}$,
which forbids the $cd=xx$ and $cd=yy$ components but allows the $cd=xy$ component,
corresponding to $\kappa_L^{xyxy;z}=\kappa_L^{xyyx;z}$.
Note that $\kappa_L^{xyxy;z}$ is also allowed by the operation $C_{4z}\mathcal{T}$.
In addition, the magnetic spin Hall effect defined by $j^{a;i}=\sigma_{b}^{a;i}E_b$ with $\sigma_{b}^{a;i}=-\sigma_{a}^{b;i}$ is forbidden by symmetry.
Therefore, for this $d$-wave altermagnet, the allowed elements of $\kappa_L^{abcd;i}$
for the $\mathcal{T}$-odd PSHE are $\kappa_L^{xyxy;z}=\kappa_L^{xyyx;z}=-\kappa_L^{yxxy;z}=-\kappa_L^{yxyx;z}$.
Since only one independent component is allowed by symmetry, we focus on $\kappa_L^{xyxy;z}$ in the following.

To proceed, we first show the band structure of this model with SOC in Fig.~\ref{FIG3}(a).
We note that the altermagnetic term $J(\cos ak_x-\cos ak_y)$ dominates the spin splitting along $\Gamma-\mathrm{X}-\mathrm{M}$,
where the spin-up and spin-down states remain nearly conserved although SOC has been included.
This can be seen from the expectation value of $\hat{\sigma}_z$, see the inset colorbar of Fig.~\ref{FIG3}(a).
In addition, the nodal line along $\mathrm{M}-\Gamma$,
protected by $[C_{2\perp}^s||\mathcal{M}_{x+y}]$ in the absence of SOC,
is slightly gapped out by a finite SOC, giving two Dirac points at $\mathrm{M}$ and $\Gamma$.
In Fig.~\ref{FIG3}(b), we present the dependence of $\kappa_L^{xyxy;z}$ on the photon frequency for two SOC strengths.
Remarkably, reducing SOC does not suppress the response;
instead, the signal becomes stronger but shifts to a lower photon frequency.
Since SOC mainly affects the states along the $\mathrm{M}-\Gamma$ path,
the peak is dominated by interband transitions between the SOC-gapped nodal-line states,
as confirmed by the $\vect{k}$-resolved quantum-geometric integral kernel in the inset of FIG.~\ref{FIG3}(b).
A smaller SOC reduces the band separation $\epsilon_{mn}$ along $\mathrm{M}-\Gamma$,
which lowers the resonant frequency through the optical selection rule $\delta(\hbar\omega-\epsilon_{mn})$ 
and simultaneously enhances the quantum-geometric integral kernel, yielding a stronger peak.

We next discuss the angular dependence of the $\mathcal{T}$-odd PSHE excited by LPL.
For LPL described by $\vect{E} (t) = E_1 (\cos\theta,\sin\theta,0)$,
the transverse spin current in a dc bias $E_0$ along the $y$ direction is
$j^{x;z}
=
\tau^2
\kappa_L^{xyxy;z}
E_0
2\text{Re}[E_{\omega}^xE_{\omega}^{y*}]
+
\tau^2
\kappa_L^{xyyx;z}
E_0
2\text{Re}[E_{\omega}^yE_{\omega}^{x*}]
=
2
\tau^2
\kappa_L^{xyxy;z}
\sin2\theta
E_0
E_1^2
=
\bar{\kappa}^{x;z}_LE_0E_1^2
$,
where we have used $\kappa_L^{xyxy;z}=\kappa_L^{xyyx;z}$ and defined
$\bar{\kappa}^{x;z}_{L}(J,\theta) = 2\tau^2 \kappa_L^{xyxy;z}\sin(2\theta)$.
The angular dependence of $\bar{\kappa}^{x;z}_L(J,\theta)$ for $J=1\,\mathrm{eV}$ and $J=-1\,\mathrm{eV}$
are shown in Fig.~\ref{FIG3}(c) and ~\ref{FIG3}(d), respectively,
where we have taken $\tau^2\kappa^{xyxy;z}_L=3\,\mathrm{nm}^2\cdot\mathrm{mA}/\mathrm{V}^3\cdot\frac{\hbar}{2e}$ in terms of Fig.~\ref{FIG3}(b).
The dependence of $j^{x;z}$ on the altermagnetic N\'eel order is manifested by the interchange of the positive and negative regions
when the altermagnetic order parameter $J$ is reversed,
demonstrating that the spin Hall current of the $\mathcal{T}$-odd PSHE satisfies
\begin{align}
j^{x;z}(J=1, \theta)=-j^{x;z}(J=-1, \theta).
\end{align}
This in turn indicates that the generated spin Hall current can serve as an optical probe of the reversal of the altermagnetic N\'eel vector.
Note that $\bar{\kappa}_L^{x;z}$ vanishes for $\theta=n\pi/2$ with $n=0, 1, 2, \cdots$
but reaches opposite extrema along the diagonal polarization directions,
as shown in Fig.~\ref{FIG3}(c-d).

\bigskip
\noindent{\textcolor{blue}{\textit{Summary}}}---In summary,
we have established a dissipationless photovoltaic spin Hall effect driven by SCV in the presence of optical excitation and a static electric field.
Circularly and linearly polarized light selectively combine SCV with the Berry curvature and quantum metric, respectively,
giving rise to $\mathcal{T}$-even and $\mathcal{T}$-odd PSHEs.
The $\mathcal{T}$-even spin Hall current can be reversed by switching the light helicity,
as demonstrated in monolayer WTe$_2$,
whereas the $\mathcal{T}$-odd response reverses with the N\'eel vector in a $d$-wave altermagnet,
providing an optical probe of altermagnetic order.
More broadly, we identify the SCV dipole as the common geometric origin of the Drude and intrinsic nonlinear spin Hall effects proposed recently,
see the Supplementary Material~\cite{sup}.
Our work therefore extends SCV from the linear magnetic spin Hall effect to
a unifying framework for dissipationless nonlinear spin Hall phenomena,
while offering new routes for optical generation and control of spin currents.

\end{document}